\documentclass[preprint,prx,amsmath,groupaddress,aps,subeqn,showpacs]{revtex4}
\pdfoutput=1
\usepackage{graphicx}

\begin{document}

\title{Controlling electron propagation on a topological insulator surface via proximity interactions}

\author{Xiaodong Li}
\affiliation{Department of Electrical and Computer Engineering, North Carolina State University, Raleigh, NC 27695-7911}


\author{Ki Wook Kim} \email{kwk@ncsu.edu}
\affiliation{Department of Electrical and Computer Engineering, North Carolina State University, Raleigh, NC 27695-7911}



\begin{abstract}
The possibility of electron beam guiding is theoretically explored on the surface of a topological insulator through the proximity interaction with a magnetic material.  The electronic band modification induced by the exchange coupling at the interface defines the path of electron propagation in analogy to the optical fiber for photons. Numerical simulations indicate the guiding efficiency much higher than that in the "waveguide" formed by an electrostatic potential barrier such as p-n junctions.  Further, the results  illustrate effective flux control and beam steering that can be realized by altering the magnetization/spin texture of the adjacent magnetic materials.   Specifically, the feasibility to switch on/off and make a large-angle turn is demonstrated under realistic conditions.  Potential implementation to logic and interconnect applications is also examined in connection with electrically controlled magnetization switching.

\end{abstract}

\maketitle
\clearpage

\section{INTRODUCTION}
Three-dimensional topological insulators (TIs) are a recently discovered quantum state of matter that is insulating in the bulk but metallic on its surface, conditioned by exceptionally strong spin-orbit interactions \cite{Qi2011,Hasan2010,Fu2007,Chen2009TI}.  The surface electrons can be described by the Dirac equation, as in graphene, leading to a linear dispersion and zero band gap (i.e., massless Dirac fermion) that is protected by the time-reversal symmetry \cite{Shan2010,Zhang2009_TI}.  In contrast to the pseudospin in graphene, the surface states of a TI are spin-polarized  with the spin direction locked perpendicular to the momentum \cite{Hsieh2009,Hsieh2009_2}, suppressing electron backscattering by nonmagnetic scatterers of arbitrary potential shape including the point defects \cite{Roushan2009,Checkelsky2009}.  These peculiar properties of TIs can be modified by breaking the very symmetry of time-reversal, leading to numerous physical phenomena of interest to such applications as spintronics and quantum computing \cite{Moore2010,Wang2012_Science_TI,Nayak2008,Yu2010}.  One fascinating example is the possibility to realize electron beam (e-beam) guiding in analogy to optical systems that was proposed initially in the graphene based structures \cite{Williams2011,Qiao2011}.


In the case of graphene, electron confinement in the guiding region can be achieved by forming electrostatic potential barriers using gate bias voltages. Depending on the carrier types, two basic guiding approaches have been investigated; namely, bipolar p-n junction guiding and unipolar fiber-optic guiding \cite{Williams2011}. The similarly chiral nature of the TI surface fermion also makes it possible application of an essentially same scheme (i.e., based on the electrostatic potential).  One crucial difference of the TI, however, is  its dependence on {\em real} spin that enables additional controllability through the interaction with adjacent magnetic  materials \cite{Wu2010,Yang2011,Mondal2010,Hammer2012}.   More specifically, the exchange coupling between the TI and a proximate (ferro- or antiferro-) magnet can break the symmetry and change the TI surface band structure.  Accordingly, the magnetization direction of the magnetic layer can strongly influence electron transmission characteristics  as evidenced by the sizable magnetoresistance predicted in a number of theoretical studies \cite{Yokoyama2010a,Kong2011,Salehi2011}.  The effective "junctions" formed by the nonuniform spin texture may thus offer an alternative means to confine, guide and modulate electron propagation on the TI surface.

The aim of this investigation is to exploit the unique advantages of the TI-magnet hybrid structures for e-beam steering with potential applications to logic and interconnects.  The ability to perform crucial operations such as turning off the flux and rounding a sharp corner is demonstrated by numerical simulations based on the finite-difference-time-domain (FDTD) and non-equilibrium Green's function (NEGF) methods \cite{Jang2010,Anantram2008}.  Finally the possibility to combine electrical control of magnetization for high functionality is discussed.


\section{ENABLING CONCEPTS}
The exchange interaction between a TI and a magnetic material is described as a perturbation added to the Hamiltonian for the low-energy TI surface electrons \cite{Zhang2009_TI,Shan2010},
\begin{equation}
H=\hbar v_{F}( {\mathbf{\sigma }}\times \mathbf{k})\cdot \mathbf{\hat{z}}+U
+\alpha\mathbf{M}\cdot {\mathbf{\sigma }},  \label{Ham}
\end{equation}
where $\mathbf{k}$ is the electron momentum, $\mathbf{\hat{z}}$ the unit vector normal to the TI surface, $v_{F}$ the Fermi velocity ($=4.28\times 10^7$ cm/s for Bi$_2$Se$_3$) \cite{Shan2010}, $U$ the electrostatic potential energy induced by the gate bias, and $ {\mathbf{\sigma }}=(\sigma_{x},\sigma _{y},\sigma _{z})$ the vector Pauli matrices for the electron spin.  The proximity coupling under discussion is described by the last term (i.e., $\alpha \mathbf{M}\cdot {\mathbf{\sigma }}$), where $\mathbf{M}= (M_x,M_y,M_z)$ denotes the magnetization in the magnetic layer and $\alpha $ is proportional to the exchange integral \cite{Yokoyama2010a,Kong2011}. For convenience, an \emph{insulating} or \emph{dielectric} ferromagnet (FM) is assumed hereinafter.  Note that an antiferromagnetic dielectric would work equally well since only the magnetization of the first layer at the interface comes into play, while a metallic counterpart could cause unintended changes in the TI electron density and, thus, is not deemed desirable.  If the magnetization of the FM is in the $z$ direction, the exchange interaction opens a band gap  in the TI surface states ($2\alpha M_z$). For the case of in-plane magnetization, the proximity effect can simply be regarded as renormalization of electron momentum for the first term in Eq.~(1); i.e., $k\mathbf{_{y}}\rightarrow k\mathbf{_{y}}+\alpha M_x/\hbar v_{F}$ and $k_{x}\rightarrow k_{x}-\alpha M_y/\hbar v_{F}$. In other words, the Dirac cone is shifted along the direction perpendicular to the magnetization.

\begin{figure}
\centering
\includegraphics[width=8.5cm]{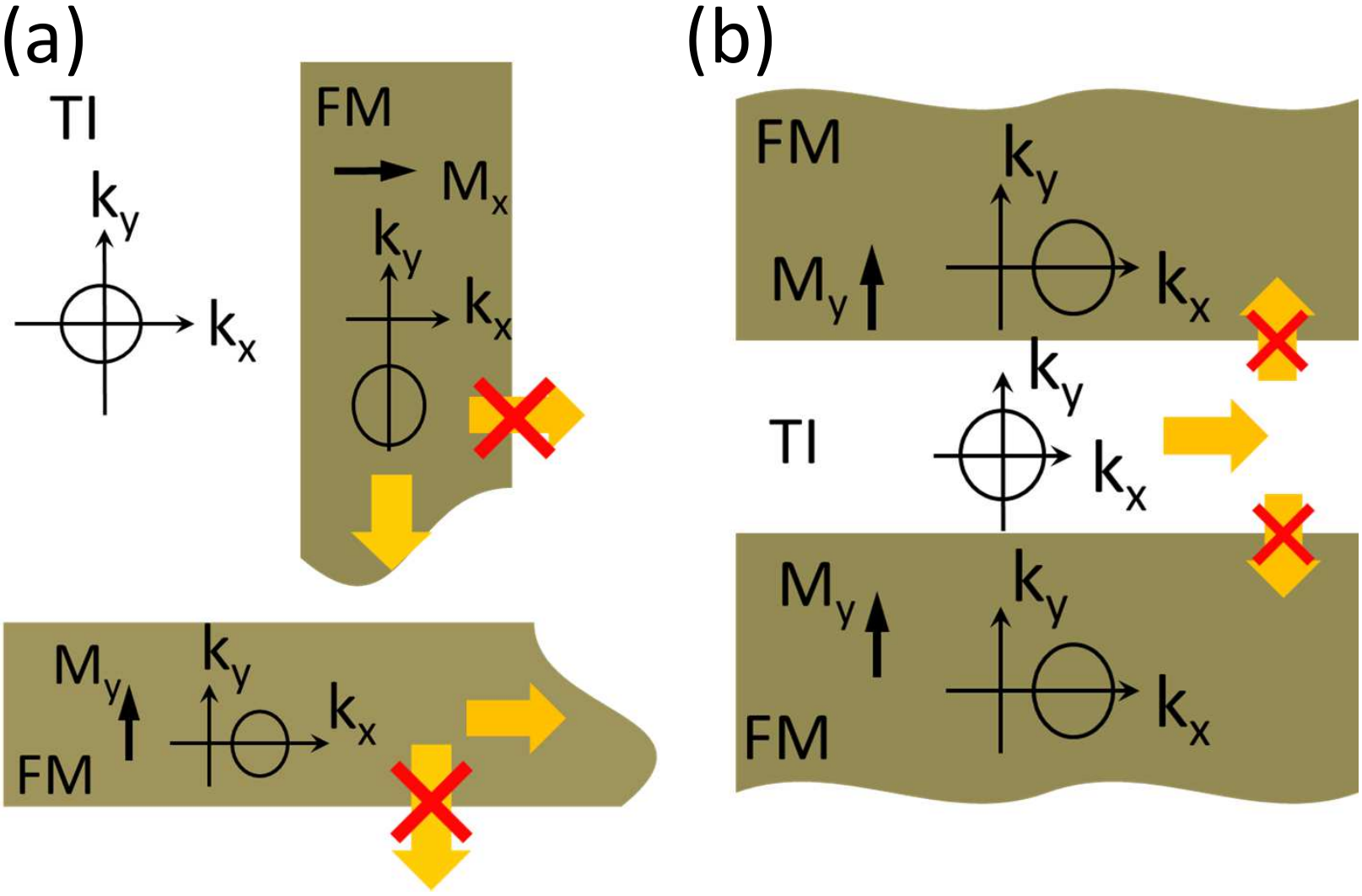}
\caption{(Color online) Proposed e-beam guiding on the surface of a TI. The path is defined by (a) a single strip of magnetic material or (b) the region between two magnetic layers.  The magnetization ${\mathbf M}$ (thin arrows) must be normal to the direction of the waveguide (i.e., e-beam propagation) indicated by the thick arrows.  The iso-energy contour of the TI surface band structure is illustrated schematically in each region.}
\label{guiding-f1}
\end{figure}

The proposed "waveguide" is defined by a single FM strip as shown in Fig.~\ref{guiding-f1}(a). In order to confine the electrons, the edges of the waveguide should be oriented perpendicular to the magnetization. For instance, the waveguide along the $y$ axis requires the top FM layer with the magnetization in the $x$ direction, and vice versa.  Electron localization inside the desired guiding region on the TI surface can be readily understood by examining the momentum conservation at the boundaries.  For the $y$-oriented waveguide (with $\mathbf{M} \parallel \mathbf{\hat{x}}$), the Dirac cone is shifted in the same direction causing mismatch of $k_y$ across the edges of the waveguide.  Since this quantity ($k_y$) must be conserved due to the invariance parallel to the interface, transmission across the boundary is prohibited and the electrons must remain inside the waveguide at least in the first-order estimation.  Adopting similar principles, the electron path can also be defined by the regions between two parallel FM strips with proper in-plane magnetization [see Fig.~\ref{guiding-f1}(b)]. In addition, the latter scheme can employ the non-zero gap of the out-of-plane magnetization for electron confinement that corresponds to the conventional semiconductor heterostructures \cite{Hammer2012}.  However, the waveguide utilizing a single FM strip is expected to be more effective as the FM strip is located directly on top of the electron path and can take advantage of spatially varying magnetization configurations simultaneously.  Accordingly, this structure [Fig.~\ref{guiding-f1}(a)] is the main focus of investigation.  Here, it is also important to point out that the term waveguide is used in a loosely defined manner as the envisioned operation does not require the fully wave nature. The key feature that enables electron confinement and guiding is simply a consequence of the induced mismatch in the band structure across the boundary.

\begin{figure}
\centering
\includegraphics[width=8.5cm]{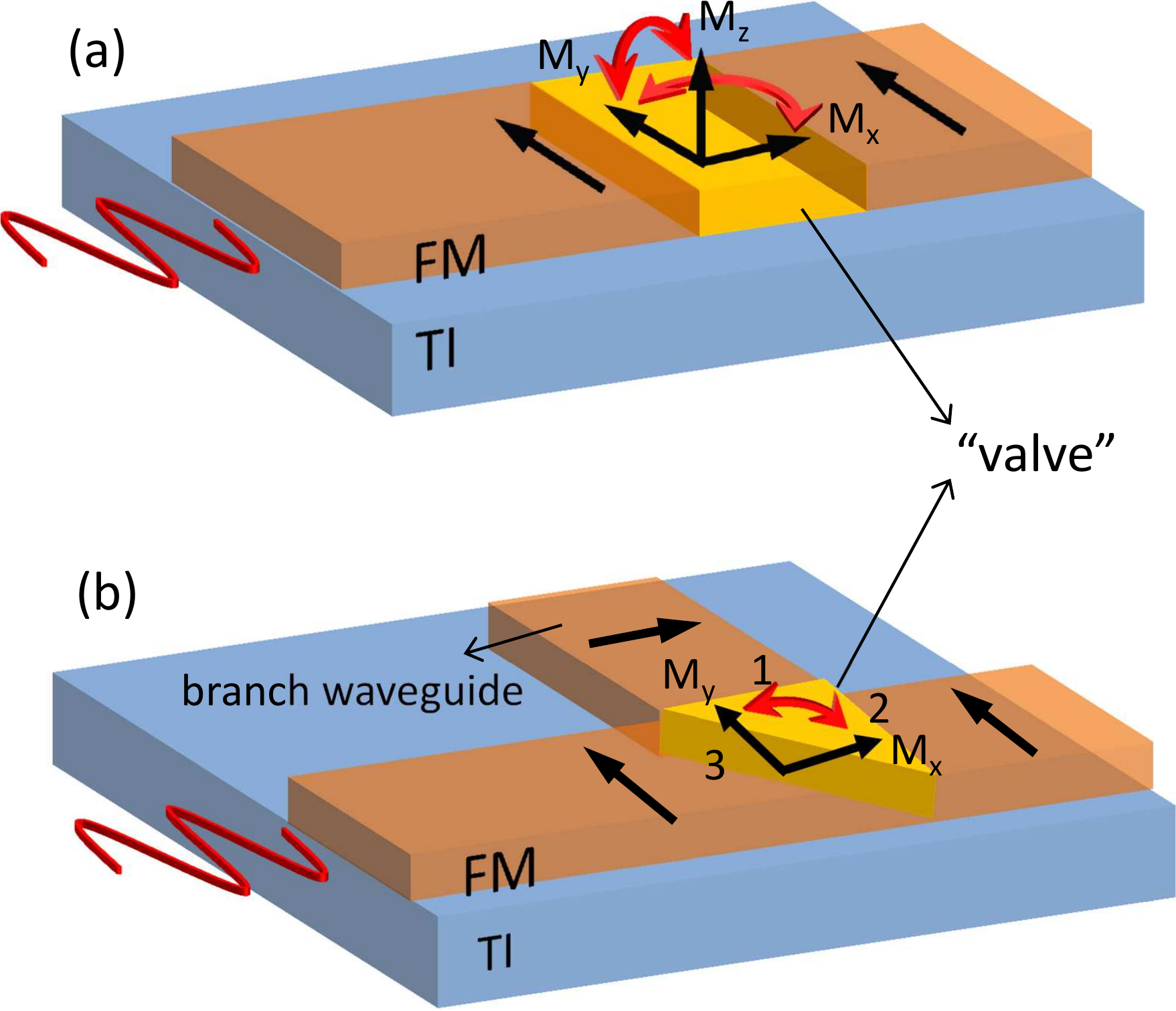}
\caption{(Color online) Device schematics for active modulation of electron propagation: (a) flux control (on/off) and (b) beam steering to a branch waveguide (a 90$^\circ$ turn).  The concepts take advantage of magnetization rotation in the control region ("valve"), either in the plane ($ M_y $ $\leftrightarrow$ $M_x$) or out of the plane ($ M_y $ $\leftrightarrow$ $M_z$).  In (b), 1,2,3 denote three boundaries, respectively.}
\label{guiding-f2}
\end{figure}

The versatility of magnetization defined e-beam guiding is further illustrated by two key extensions of the concept whose functions are essential for active modulation of electron propagation in the waveguide; namely, flux control and beam steering as indicated in Figs.~\ref{guiding-f2}(a) and (b), respectively.   Figure~\ref{guiding-f2}(a) considers a case where the magnetization in a segment of the FM layer can switch its direction by 90$^\circ$.  The waveguide is in the "on" state once the control magnetization (i.e., the "valve") is aligned with that of the main channel.  For the "off" state, on the other hand, the magnetization is rotated by 90$^\circ$, either in the plane or out of the plane.  The in-plane switch (i.e., $M_y\rightarrow M_x$) induces displacement of the Dirac cone normal to the waveguide while the out-of-plane switch ($M_y\rightarrow M_z$) opens a band gap, both of which block electron transmission. Figure~\ref{guiding-f2}(b) shows a schematic, where the main waveguide is connected with a second branch along the $y$ direction through a triangular shaped control region.  By aligning the "valve" with either of the channels, the electrons can be made propagating along the main waveguide or diverted into the branch after rounding a 90$^\circ$ corner.  The triangular shape for the control region is critical for steering the e-beam.  When the valve is aligned with the main waveguide magnetization, the band mismatch between the branch and the main waveguide does not allow conservation of the lateral momentum across the boundary [see the edge marked "1" in Fig.~\ref{guiding-f2}(b)], maintaining the electrons in their initial path.  Once this is switched parallel to the magnetization of the branch waveguide, the boundary is formed at the hypotenuse of the triangular region (i.e., edge 3).  Since the amount of Dirac cone shift projected on the boundary direction is the same on both sides of edge 3, the momentum conservation rule can be easily satisfied along this boundary.  Thus, the electrons can transmit through the interface into the control region and, then, proceed to the branch waveguide.  At the same time, edge 2 blocks electron propagation to the main waveguide.  The net effect completes a $90^\circ$ turn by the propagating e-beam.

\section{NUMERICAL APPROACH}
To clearly demonstrate the electron guiding phenomena, we adopt the FDTD method and solve numerically the time-dependent Dirac equation including the proximity effect. This method has been used successfully to investigate the electronic behavior with quasi-optical dynamics such as Klein tunneling and the Goos-H\"{a}nchen effect in graphene \cite{Jang2010}. The FDTD simulation is performed in a 500 nm$\times$300 nm region.  The waveguide of 100 nm in width is placed in the middle along the $x$ direction. The mesh size is set at 1 nm$\times$1 nm.  To consider excitation of electrons propagating in all directions, a point source of continuous sinusoidal wave is used.  The simulation region is assumed to be surrounded by the boundary layers that can absorb all electrons going out of the simulation region with little reflection.

While the FDTD simulation gives a direct and intuitive picture of electron guiding as well as the basic operations, it is also known to experience difficulties in obtaining numerically reliable data under low transmission due to the mesh size dependence and reflection error from the boundary \cite{Jang2010,Taflove2005}. Accordingly, a complementary approach based on the NEGF formalism is adopted for accurate evaluation of the transmission function (a local property) and, subsequently, the on/off ratio for the electron flux \cite{Anantram2008}.  The calculation is carried out for a one-dimensional case, assuming that the width of the waveguide is sufficiently large with a negligible impact on the final outcome.  In this formalism, the retarded Green's function is defined as \cite{Anantram2008}
\begin{equation}
G(E,k_y)=[E-H+i\eta -\Sigma _1(E,k_y)-\Sigma _2(E,k_y)]^{-1},
\end{equation}
where $H$ is the Hamiltonian defined in Eq.~(1) including the proximity effects, $\eta$ symbolizes an infinitesimally small positive number, and $\Sigma _1$ and $\Sigma _2$ are the self-energies of the semi-infinite leads on the left and right, calculated by using the Sancho-Rubio iterative method \cite{Sancho1984,Nardelli1999}. The transmission function can be obtained as
\begin{equation}
T(E,k_y)= \mathrm{Tr}[\Gamma _1(E,k_y)G(E,k_y)\Gamma _2(E,k_y)G(E,ky)^+],
\end{equation}
where $\Gamma _{1,2}=i(\Sigma _{1,2}-\Sigma _{1,2}^+)$. With the transmission function, the conductance for the electron flux ($G^C$) can be calculated by using the Landauer formalism:
\begin{equation}
G^C =\frac{e^2}{\hbar}\iint\frac{dE}{2\pi} \frac{dk_y}{2\pi} ~T(E,k_y)\left[-\frac{\partial f_0(E)}{\partial E}\right]_{E=E_F},
\end{equation}
where $e$ is the electron charge, $f_0$ the Fermi-Dirac distribution function, and $E_F$ the Fermi level at the contacts.  The Dirac point serves as the reference for energy (i.e., zero).

\section{RESULTS AND DISCUSSION}
\begin{figure}
\centering
\includegraphics[width=8.5cm]{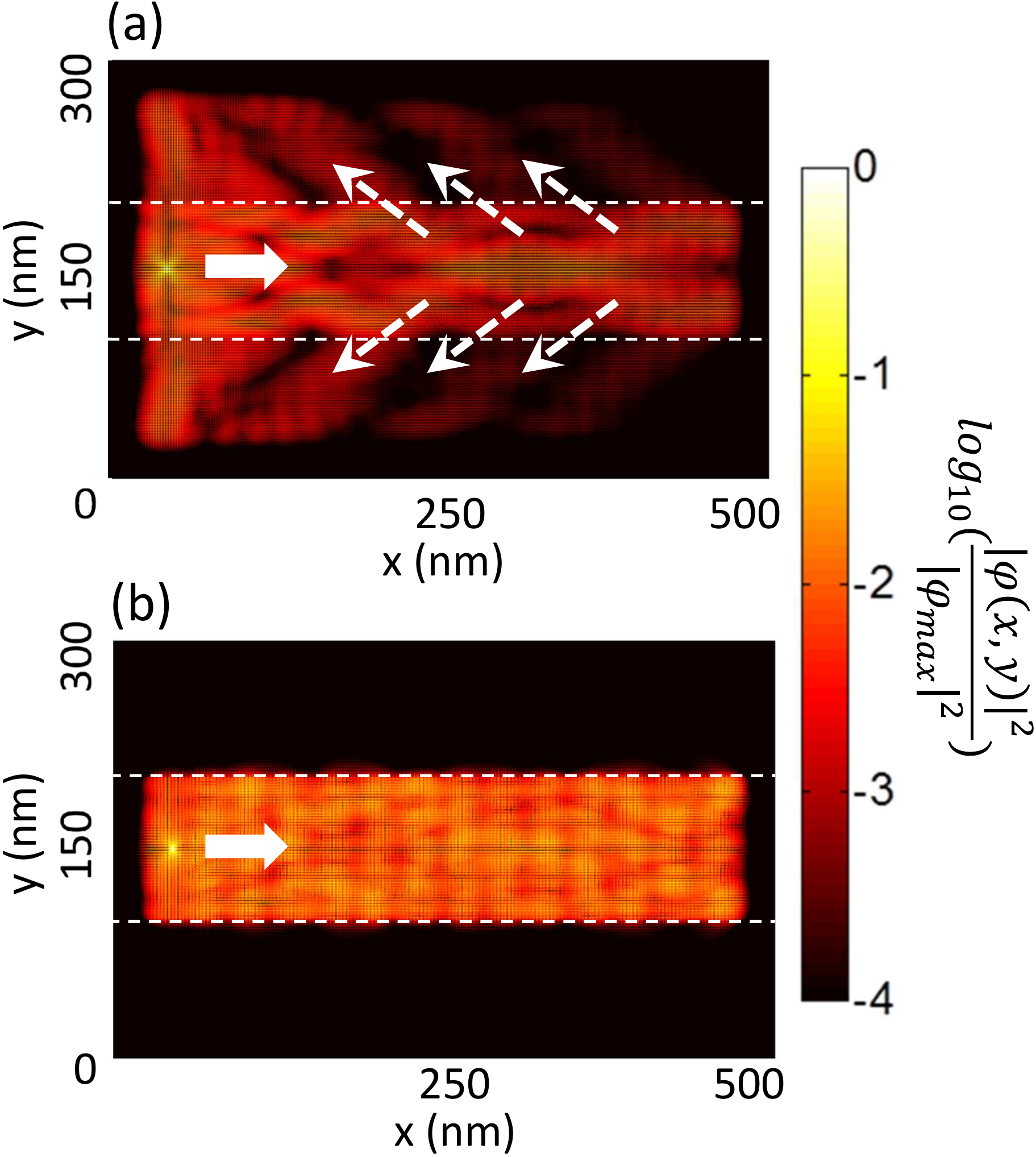}
\caption{(Color online) FDTD calculation of electron probability density function in a waveguide relying on (a) the electrostatic potential barrier or (b) the proximity effect with a FM.  Both the electrostatic barrier height and the exchange interaction energy are set at 100 meV.  The injected electron energy is 50 meV. The probability density is normalized to the maximum value. The width of the waveguide is 100 nm.}
\label{guiding-f3}
\end{figure}

Figure~\ref{guiding-f3} shows the FDTD simulation results in two different waveguides relying on (a) the potential barrier induced by a gate bias or (b) the proximity effect through the exchange interaction with a FM, respectively.  To be comparable, both the bias induced electrostatic barrier height and the exchange interaction energy, $E_{ex}=|\alpha\mathbf{M}|$, are set at 100 meV.   The injected electron energy is chosen to be one half of this value (i.e., 50 meV). A point electron source is assumed near the left end of the waveguide in the calculation; the e-beam travels along the $x$ axis due to the confinement of the effective barrier. The normalized electron probability density function is plotted once the system reaches a stable state after a sufficiently long simulation time.  Note that the color bar is in the logarithmic scale. In the case of Fig.~\ref{guiding-f3}(a), strong leakage is clearly visible outside the waveguide indicating a poor guiding efficiency.  It is also interesting to see that the leaked e-beam propagates in the backward direction (i.e., $-x$) resembling optical refraction in the Veselago lens.  When the electron crosses the edge of the waveguide, it transfers from the conduction band to the valence band, which shows a pattern similar to the graphene p-n junction \cite{Cheianov2007}.  Since the barrier potential is chosen to be just twice the electron energy, the equivalent index  is $-1$ in this  region.  The fundamental issue of the bias induced guiding is that the electrostatic potential barrier cannot completely confine electrons due to the absence of band gap. Our simulation indicates that the leakage indeed remains quite prominent even with different choices of barrier heights in concert with the theoretical understanding. By contrast, the band modification via the the proximity interaction between the TI and FM can confine electron propagation  very effectively; the amplitude of the probability density function outside the waveguide is several orders of magnitude smaller as illustrated in Fig.~\ref{guiding-f3}(b).  When the electron encounters the boundary, it experiences the absence of an appropriate state outside the waveguide owing to the mismatch of the bands. Thus, the e-beam is reflected back to the waveguide, producing an excellent guiding efficiency.

The simulation also indicates that the leakage increases drastically even in the case of band modification if electrons of higher energy are injected.  As the radius of the iso-energy contour becomes larger (i.e., large $k$), it is progressively more difficult to completely separate the bands inside/outside the waveguide via the shift of Dirac cones.  Once an overlap occurs between them, the condition of lateral momentum conservation can be satisfied in this region of the momentum space and the electrons are allowed to propagate readily across the boundary.  As such, only the electrons within the energy range of $\pm E_{ex}/2$ astride the Dirac point can be well confined in the waveguide since the bands inside/outside the waveguide do not overlap.  This characteristic enables the waveguide to work as an energy filter.  A consequence is that the electron distribution inside a sufficiently long waveguide could be limited to  $ - E_{ex}/2 \lesssim E \lesssim + E_{ex}/2$ even when the injected is broader; the high energy tail leaks out very quickly.

\begin{figure}
\centering
\includegraphics[width=8.5cm]{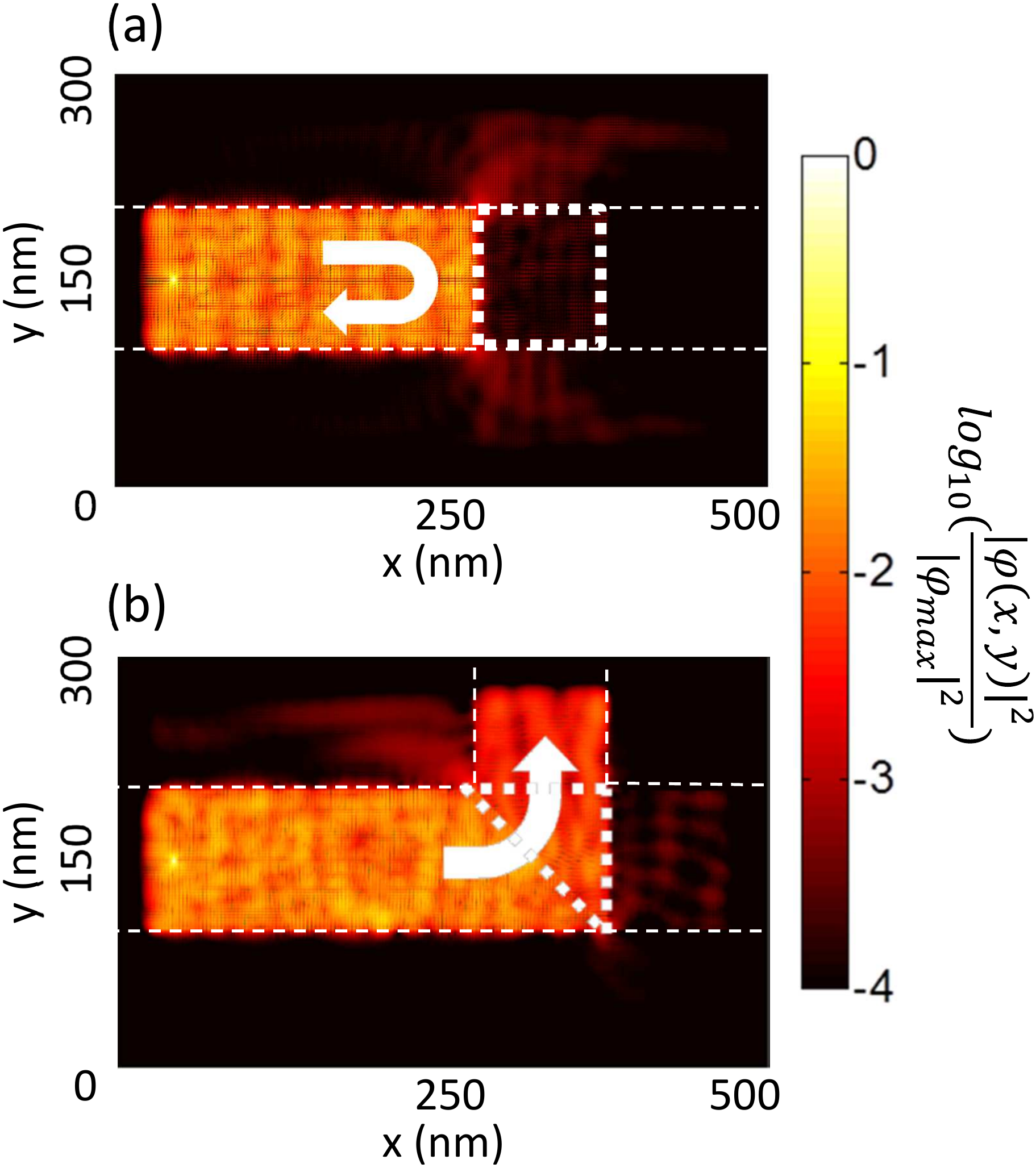}
\caption{(Color online) FDTD calculation of electron probability density function demonstrating (a) flux control (the off state) and (b) beam steering to a branch waveguide (a 90$^\circ$ turn). The parameters are the same as those specified in Fig.~\ref{guiding-f3}.}
\label{guiding-f4}
\end{figure}

Figures~\ref{guiding-f4}(a) and (b) provide the calculation results demonstrating flux control and beam steering, respectively.  The "off" state (i.e., low or no flux/current) is realized by turning the (rectangular shaped) valve by 90$^\circ$ in the plane, from $M_y$ to $M_x$. The length of the control region is chosen to be 100 nm, the same as the waveguide width.  As shown, the electrons are effectively reflected back by the band mismatch at the interface; only a very tiny fraction either transmits through the barrier or leaks outside the waveguide region.  For e-beam steering in Fig.~\ref{guiding-f4}(b) with a triangular shaped control, the loss is relatively larger as a small portion  continues its passage in the main waveguide instead of turning the corner as desired.  In order to curtail this leakage, an additional barrier [such as that used in Fig.~\ref{guiding-f4}(a)] may be necessary.  Note that the operating principles discussed above are expected to remain valid even in the presence of electron scattering events in the channel.  As they are determined by transmission/reflection right at the boundaries (i.e., local properties),  ballistic transport in the entire device is not required for e-beam guiding/steering and flux control in the TI-magnet hybrid structure.

\begin{figure}
\centering
\includegraphics[width=8.5cm]{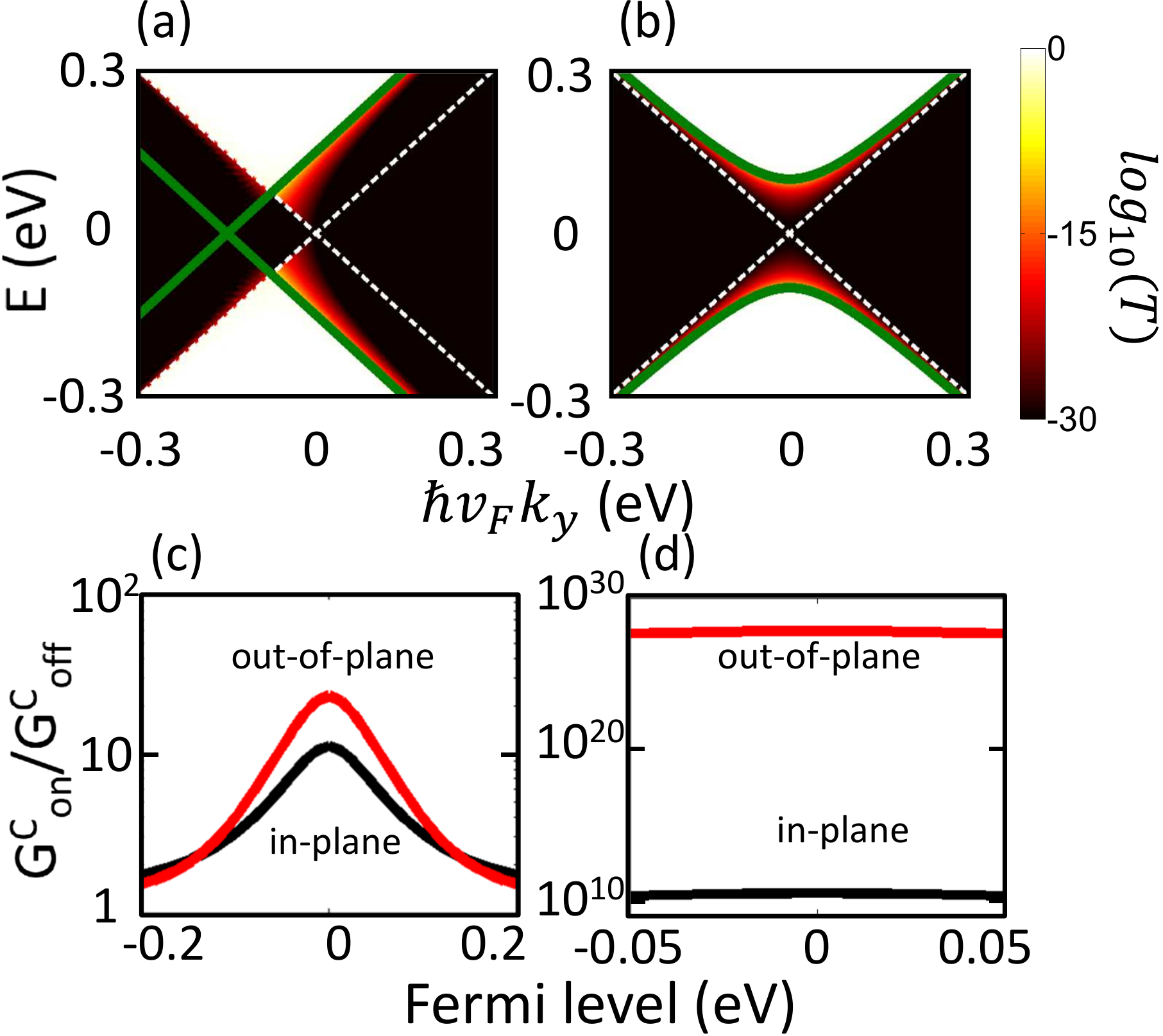}
\caption{(Color online) NEGF simulation results for the flux control operation. (a,b) Calculated transmission coefficient as a function of electron momentum $k_y$ and energy $E$ for the in-plane and the out-of-plane off states, respectively. The dashed lines depicts the Dirac cone in the waveguide, while the thick solid lines indicate the Dirac cone in the control valve region. (c,d) Conductance on/off ratio versus Fermi level at 300 K with the thermal electron distribution and the distribution truncated outside the energy range [$-E_{ex}/2,+E_{ex}/2$], respectively.  Case (d) is to emulate the leakage of high energy electrons while propagating in the waveguide region.}
\label{guiding-f5}
\end{figure}

The issue of flux control is further examined by the NEGF method for a more quantitative analysis. Figures~\ref{guiding-f5}(a) and (b) illustrate the calculated transmission probability $T(E,k_y)$ as a function of electron momentum $k_y$ and energy $E$ for the off states, where the control magnetization is switched to the $x$ or the $z$ direction, respectively [see Fig.~\ref{guiding-f2}(a) for reference].  In each case, the dashed lines denote the Dirac cone in the waveguide, whereas the thick solid lines indicate the modified dispersion in the control region. The overlap of these band structures determines the states with nonzero transmission. As evident from the figures, the mismatch of the band produces an equivalent band gap for the incoming electrons in the waveguide, which can effectively prohibit electron propagation in a specific range of energy.  In fact, a properly tuned monochromatic beam (that is analogous to the  injection condition of FDTD calculations) would experience a precipitous drop of several orders of magnitude in the transmitted flux to nearly total reflection as the control valve switches to the off state.  Further, the forbidden energy range is directly dependent on the exchange energy between the TI and FM layers $-$ a large exchange energy is preferred in order to block the passage of electrons in a wide energy spectrum.  A recent experiment reported that a band gap up to $ \sim 100$ meV can be generated in the surface states of Bi$_2$Se$_3$ by doping magnetic impurities (such as Cr) \cite{Kou2012}, clearly suggesting $\alpha M$ of the same order as that used in the current study.  At the same time, it is plausible to anticipate an even higher exchange coupling energy in a fully optimized system.

Along with the transmission probability, it is also revealing to examine the channel on/off ratio from the perspective of magnetoresistance. 
Utilizing the Landauer formalism as discussed earlier, Fig.~\ref{guiding-f5}(c) shows the conductance on/off ratio estimated at room temperature for the in-plane and out-of-plane switches (with the control valve length of 100 nm).  It is clear from the figure that the obtained values are much smaller than those expected from the transmission probability or Fig.~\ref{guiding-f4}(a), with the maximum on/off ratio reaching only approximately 23.  This can be attributed to thermal broadening in Eq.~(4); the injected electrons have a substantial high energy tail that cannot be blocked completely by the the off states.

One potentially crucial point to note, however, is that the above treatment does not consider the energy filtering effect in the beam guiding region. Of the thermal electrons injected from the contacts, the high-energy population leaks out of the channel with exponentially larger probabilities.  Subsequently, the guided e-beam is expected to have a distribution with a suppressed tail in the realistic channels that could drastically reduce the off-state current, while its impact on the on state is relatively minor.  To emulate this condition, we compute the channel conductance by considering the electrons only within the range of high guiding efficiency ($ - E_{ex}/2 \leq E \leq + E_{ex}/2$). As expected, the results plotted in Fig.~\ref{guiding-f5}(d) yield a much higher on/off ratio even at room temperature $-$ approx. 10$^{27}$ and 10$^{10}$ for the out-of-plane and the in-plane switch, respectively.  Similarly, one may also ensure a narrow spectral range for the e-beam by taking advantage of tunnel injection.  In addition, another strategy to curtail the leakage (thus, enhancing the on/off ration) is to implement a second valve in series that can engineer band modification different from the first.  The presence of two barriers in series can provide an extended energy range with little or no transmission, making the structure less susceptible to thermal broadening/noise.  Calculation of the numerically accurate on/off ratio in the realistic structures requires explicit consideration of such effect as finite channel width and scattering, which is beyond the scope of the current investigation.

\begin{figure}
\centering
\includegraphics[width=8.5cm]{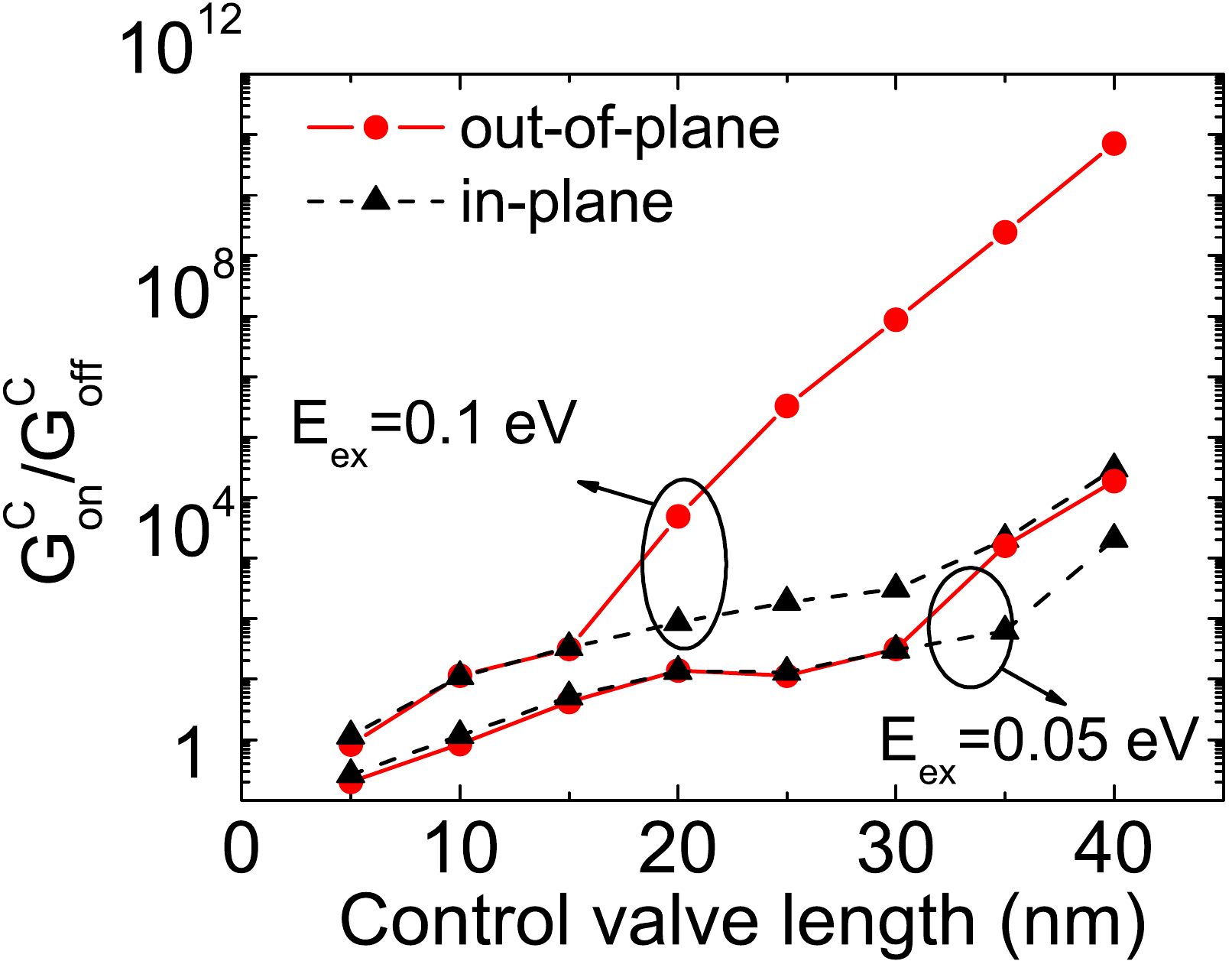}
\caption{(Color online) Conductance on/off ratio versus the length of the control region for two different exchange energies.  Both the in-plane and out-of-plane off states are considered.  The calculation is performed with the Fermi level at the Dirac point ($E=0$ eV) at 300 K by considering the thermal distribution only in the energy range [$-E_{ex}/2,+E_{ex}/2$]. See also Fig.~\ref{guiding-f5}(d).}
\label{guiding-f6}
\end{figure}

Figure~\ref{guiding-f6} illustrates the dependence of the conductance ratio on the length of the control region ("valve") at room temperature, considering only the electrons  in [$-E_{ex}/2,+E_{ex}/2$]. The Fermi level is assumed to be at 0 eV (i.e., the Dirac point).  Two values of exchange energies are used, 0.1 eV and 0.05 eV, respectively.  When the size of the control region decreases, the on/off ratio reduces drastically due to the rapid increase of the off-state conductance in accord with a shorter tunnel barrier width. On the other hand, a large exchange energy can generate a higher tunnel barrier leading to a smaller off current and a larger on/off ratio.  Of the two off states, the out-of-plane switch appears to provide a larger on/off contrast particularly when the length of the valve increases.  This is because the barrier height induced in the out-of-plane switch is two times that of the in-plane mismatch (i.e., $E_{ex}$ vs. $E_{ex}/2$).  Thus, the turn-off operation of the out-of-plane is more efficient than the in-plane switch; the difference is more prominent with thicker barriers.

\begin{figure}
\centering
\includegraphics[width=8.5cm]{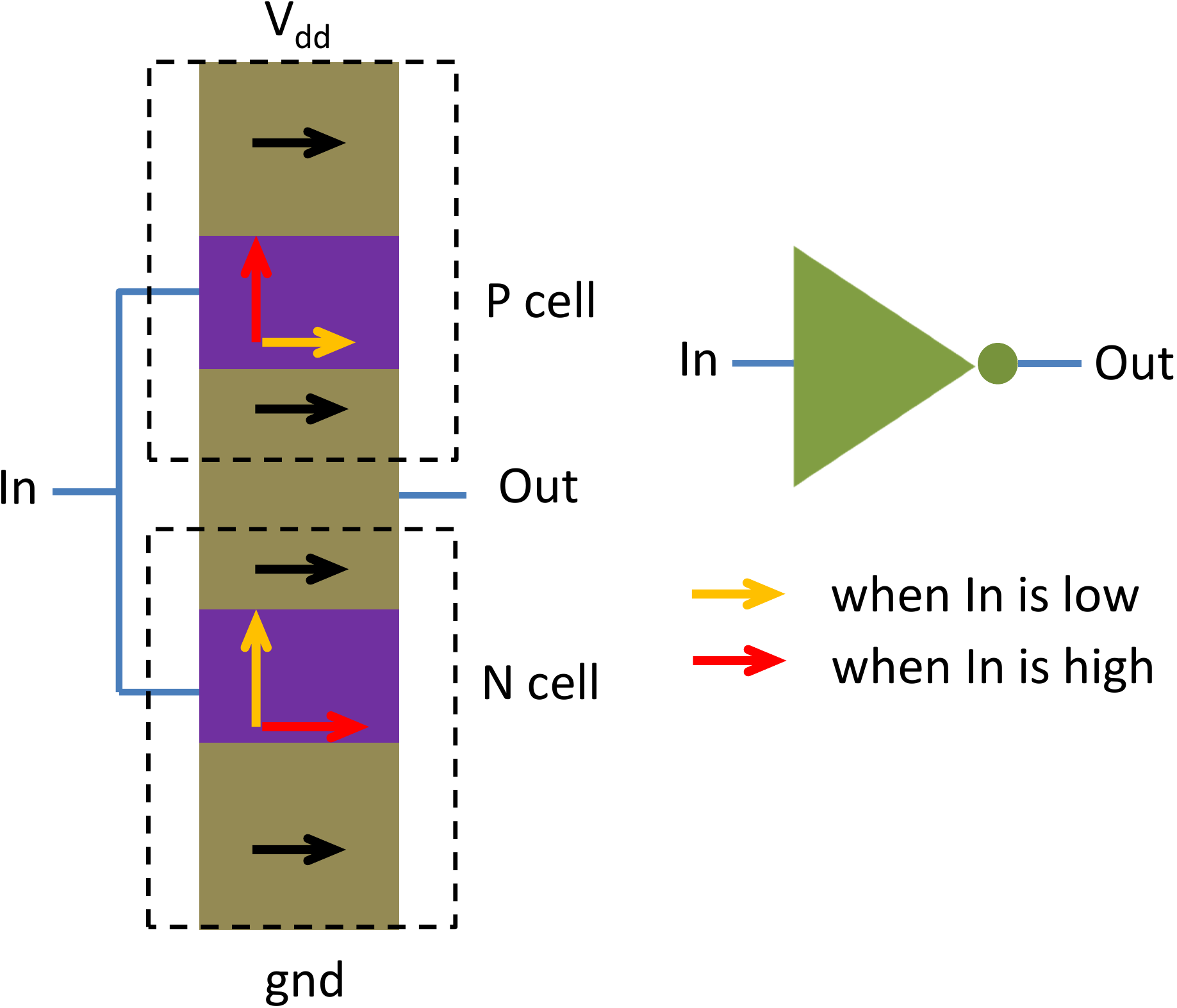}
\caption{(Color online) Potential applications of electron guiding in logic circuits: inverter implementation using the electron waveguide. The yellow arrow denotes the magnetization direction when the input (In) is in the "low" state, whereas the red arrow corresponds to the magnetization when the input is "high".}
\label{guiding-f7}
\end{figure}

The basic principles of e-beam guiding discussed above can be extended to devices with a more complex functionality.  Clearly, a properly operating control valve is critical, i.e., to switch the magnetization by $90^\circ$ either in plane or out of plane. This can be realized by using the magnetoelectric effect or multiferroic materials/structures \cite{Wu2011,Moubah2013,Chu2008,Ramesh2007,Semenov2012}.  Here, we simply assume that the direction of magnetization can be manipulated by the gate voltage instead of going into details of the methods. Then, a rudimentary transistor-like function can be imagined by utilizing the flux controlling ability.  In fact, both "n-cell" and "p-cell" structures can be discussed by varying the initial magnetization of the control valve to be aligned either perpendicular or parallel to that of the waveguide region, respectively.  With a sufficiently high gate bias (causing a $90^\circ$ switch), the former turns from the off to the on state whereas the latter does the opposite.  Then, an inverter can be directly built by fabricating two "complementary" control gates separately on one waveguide strip, as shown in Fig.~\ref{guiding-f7}.  Other logic gates, such as NAND and NOR, can be constructed similarly.

\begin{figure}
\centering
\includegraphics[width=8.5cm]{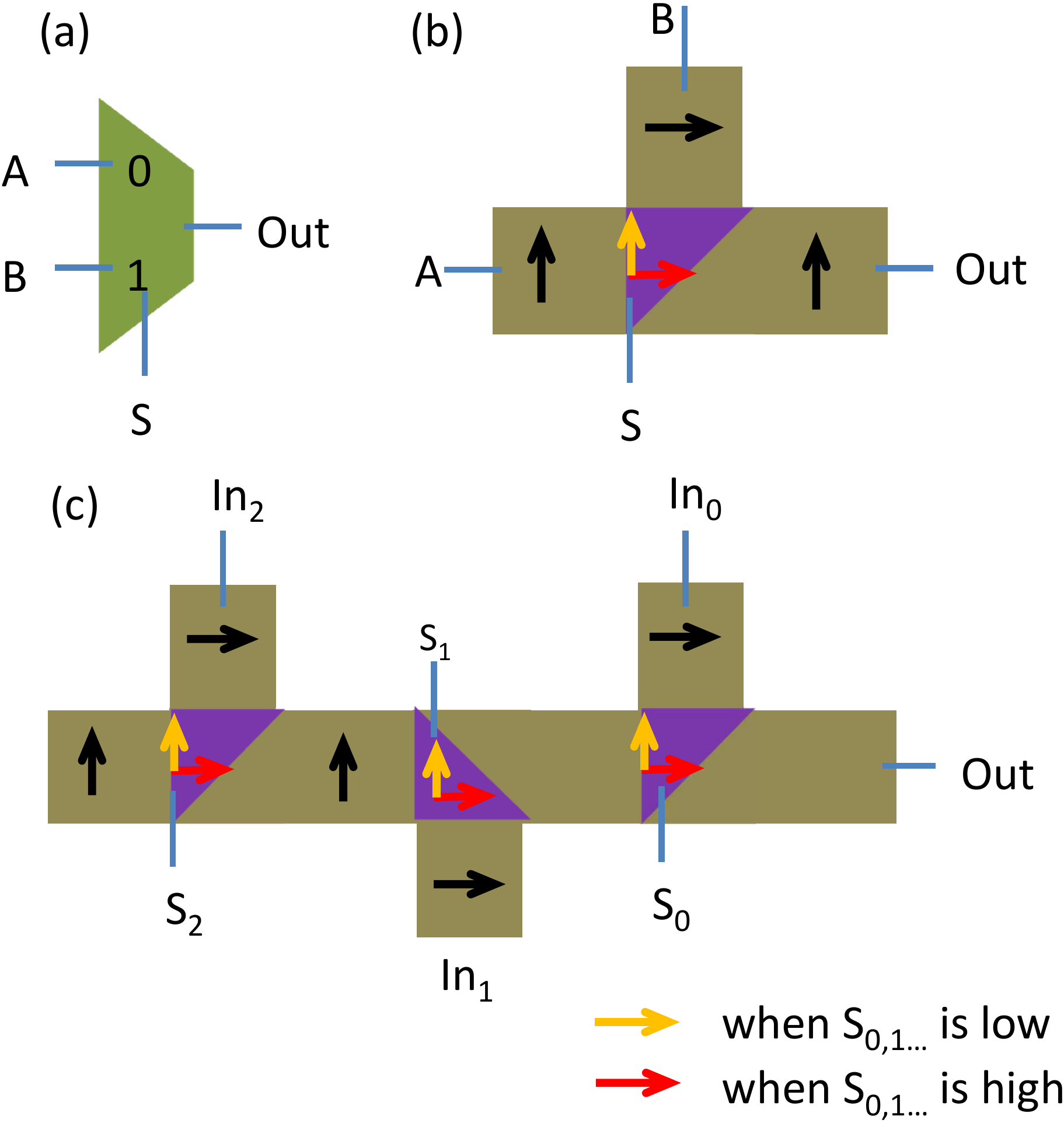}
\caption{(Color online) Potential applications of electron guiding in interconnect circuits: (a) Symbol for multiplexer; (b) multiplexer implementation using the proposed electron waveguide; (c) merged multiplexer and arbiter. The yellow arrow denotes the magnetization direction when the selection signal is low, whereas the red arrow is the case when the selection signal is high.}
\label{guiding-f8}
\end{figure}

As for the steering operation, the waveguide can be directly used as circuit interconnect that can dynamically select the signal paths.  Particularly, it can realize the multiplexing/demultiplexing function as indicated in Figs.~\ref{guiding-f8}(a) and (b).  By applying the gate voltage in the triangular control valve, the main waveguide can output the signal from either A or B. Moreover, it is a natural combination of arbitration and multiplexing if multiple branches of waveguides are connected to a single main channel as in Fig.~\ref{guiding-f8}(c).  Multiple input signals come through the branch waveguides separately, while the output signal goes out from the main waveguide. The branch waveguides near the output have higher priorities than those farther upstream.  For instance, signal ${\mathrm {In_0}}$  has the highest priority in Fig.~\ref{guiding-f8}(c); once ${\mathrm {S_0}}$ turns high, the control valve sends ${\mathrm {In_0}}$ to the output and simultaneously blocks all other signals ${\mathrm {In_1}}$, ${\mathrm {In_2}}$, ..., etc.

\section{SUMMARY}
The electron guiding phenomena are exploited on the surface of a TI through the exchange interaction with a proximate magnetic material.  Following a FDTD simulation, the guiding efficiency is found to be significantly higher than the earlier approaches based on the electrostatic potential barriers.  In particular, the leakage outside the waveguide appears to be several orders of magnitude smaller.  The theoretical analysis also demonstrates the feasibility of two essential operations  for active modulation of electron propagation (namely, flux control and beam steering) by utilizing the possibility of electrically controlled magnetization switching.  A numerical investigation based on the NEGF method further validates the concept under realistic conditions at room temperature. Finally, a number of potential electronic applications are proposed including a complementary logic, interconnect, and multiplexer/demultiplexer.  Due to the operating principles that are determined by local transmission probabilities, the envisioned devices are not expected to be limited by the requirement of ballistic transport or fully wave nature.

\begin{acknowledgments}

This work was supported, in part, by SRC/NRI SWAN, MARCO/DARPA STARnet FAME, and the US Army Research Office.

\end{acknowledgments}


\clearpage

\clearpage

\clearpage

\clearpage

\clearpage

\clearpage

\clearpage

\clearpage

\end{document}